\documentclass{article}
\usepackage[utf8]{inputenc}
\usepackage[margin=1in]{geometry}
\usepackage{bbm}
\usepackage[T1]{fontenc}
\usepackage{etoolbox}

\makeatletter
\patchcmd{\maketitle}{\@fnsymbol}{\@alph}{}{}  % Footnote numbers from symbols to small letters
\makeatother
\usepackage[titletoc,title]{appendix}
\usepackage{amsmath,amsfonts,amssymb,mathtools}
\usepackage{systeme}
\usepackage{graphicx,float}
\usepackage[ruled,vlined]{algorithm2e}
\usepackage{algorithmic}
\usepackage{authblk}
\usepackage{pdfpages}
\usepackage{enumitem}
\usepackage{hyperref}
\usepackage{xcolor}
\usepackage{amsmath}
\usepackage{url}
\usepackage{tocbibind}

\title{Pandemic Data Quality Modelling: A Bayesian Approach}
\author{Luisa Ferrari\thanks{Department of Statistical Sciences "Paolo Fortunati", University of Bologna}, Giancarlo Manzi\thanks{Department of Economics, Management and Quantitative Methods, University of Milan}, Alessandra Micheletti\thanks{Department of Environmental Science and Policy, University of Milan}, Federica Nicolussi\thanks{MOX, Dipartimento di Matematica, Politecnico di Milano, Piazza Leonardo da Vinci 32, 20133, Milano, Italy}, Silvia Salini\thanks{Department of Economics, Management and Quantitative Methods, University of Milan}}

\date{}
\begin{document}

\maketitle

\begin{abstract}
When pandemics like COVID-19 spread around the world, the rapidly evolving situation compels officials and executives to take prompt decisions and adapt policies depending on the current state of the disease. In this context, it is crucial for policymakers to have always a firm grasp on what is the current state of the pandemic, and to envision how the number of infections and possible deaths is going to evolve over the next weeks. However, as in many other situations involving compulsory registration of sensitive data from multiple collectors, cases might be reported with errors, often with delays deferring an up-to-date view of the state of things. Errors in collecting new cases affect the overall mortality, resulting in excess deaths reported by official statistics only months later. In this paper, we provide tools for evaluating the quality of pandemic mortality data. We accomplish this through a Bayesian approach accounting for the excess mortality pandemics might bring with respect to the normal level of mortality in the population.
\end{abstract}

Keywords: Pandemics, Bayesian analysis, variance models, time-space models

\section{Introduction}
Historically, human populations have always dealt with major outbreaks of infectious diseases and effective reaction to these outbreaks has always been sought. The numerous black plague outbreaks in Europe in medieval and post-medieval times, the cholera pandemic in the late nineteen- early twentieth century, and the Spanish flu in the early twentieth century are only a few examples in modern history.
However, the COVID-19 pandemic, which brought the world close to a halt in 2020 and 2021 and has killed almost seven million people as of early 2023, has brought to light a new reality of a global pandemic never experienced before.
Worldwide governments initially underscored its urgency and their healthcare systems were quickly overwhelmed in a tsunami-like fashion. Nearly all of the affected countries progressively implemented measures to slow down the spread of the virus, ranging from recommended social distancing to almost complete lockdowns of social and economic activity. 
These measures eventually proved to be effective, allowing 
numerous states to relax restrictions, in an attempt to gradually return to normality. At the same time, with the threat posed by the virus still looming,
decision-makers were forced to strike a balance between epidemiological risk and allowance of socioeconomic activity. 
In this type of context, surveillance of the number of new infections, mortality monitoring and quantification of the effects of social distancing  became increasingly important \cite{Colombo2020, Kantner2020, Wu2020}, particularly so on a regional level. Given the local nature of the phenomenon, such a regional view appears to be of crucial importance. One of the difficulties lies in the fact that exact numbers of new infections, reported deaths and recoveries from the disease are often available only with a certain probability to be later corrected or are reported with a delay of - sometimes - several days. Moreover, there are discrepancies among data reported at different levels (regional, provincial, etc.) sometimes crucially causing wrong responses to face the pandemic.

Another important aspect to keep in mind is that the deaths due to COVID-19 nationwide are underestimated by official data because deaths that have not been tested are not counted. In addition, the pandemic had an indirect effect on mortality, preventing timely treatment for other diseases or limiting preventive examinations that could have anticipated critical situations. For all these reasons, official data on COVID-19 deaths in some countries were fairly inadequate to measure the effect of the pandemic \cite{Collaborators}. In aid of this, the World Health Organization (WHO) suggests studying excess mortality for assessing the death burden (both direct and indirect) of COVID-19. Excess mortality  refers to the number of deaths from all causes during the pandemic more than what we would have expected under “normal” conditions, which is a valid measure of the total effect of the COVID-19 pandemic \cite{Beaney2020}. Worldwide studies on excess mortality are deepened as reported in \cite{Msemburi2023} and \cite{Shang2022}.
Multiple model-based mortality estimates have been proposed in the literature, \cite{Beaney2020}, \cite{Blangiardo2020}, \cite{Maruotti2022}, \cite{Michelozzi2020}, and \cite{Ceccarelli2022} at national, regional and also county level. Several studies have shown a critical increase in excess mortality for higher age groups and for males, see for instance \cite{Gibertoni2021}.

Excluding COVID-19, the ongoing global disease outbreaks reported by the WHO between 7/4/2022 and 7/4/2023 with at least one death are those listed in Table \ref{Tab1}. Most of the diseases are endemic, especially in Africa and Asia. Diseases first reported in the last 25 years are, apart from Sars-Cov2, Mers-Cov (first reported in 2012), the Nipha virus (first reported in 1999) and a new type of hepatitis in children (first reported in 2022). Of the 16 diseases in an outbreak phase, 3 were first detected in the 18th century or before, 3 in the 19th century, 8 in the 20th century and 2 in the first 23 years of the 21st century. 

\begin{table}[h]
\label{Tab1}
\caption{Disease outbreaks with at least 1 death reported by WHO. 7/4/2022 – 7/4/2023 (Source: Authors' calculations from WHO data)}
\begin{tabular}{l|l|l|l|l}
\hline 
\hline
\textbf{Affected country}    & \textbf{Last} & \textbf{Virus (presumed year of}                                          & \textbf{Case} & \textbf{Deaths} \\
\textbf{}    & \textbf{news} & \textbf{first reporting)} & \textbf{fatality} & \textbf{per day} \\
\textbf{}    & \textbf{update} & \textbf{} & \textbf{ratio} & \textbf{} \\
\hline\hline
Saudi Arabia                 & 7/4/2022                  & MERS-Cov (2012)                                                & 66.67\%                      & 0.02                    \\
Several countries            & 12/7/2022                 & Severe acute hepatitis of unknown  & 2.18\%                       & 0.23                    \\
           &                  & aetiology in children (2022) &                        &                    \\
Malawi                       & 9/2/2022                  & Cholera (19th century)                                                 & 3.28\%                       & 3.59                    \\
Dem. Rep. of Congo & 28/4/2022                 & Ebola (1976)                                                  & 100.00\%                     & 0.02                    \\
Australia                    & 28/4/2022                 & Japanese encephalitis (1871)                                   & 12.00\%                      & 0.02                    \\
Qatar                        & 12/5/2022                 & MERS-Cov (2012)                                               & 50.00\%                      & 0.002                   \\
Cameroon                     & 16/5/2022                 & Cholera (19th century)                                                & 2.01\%                       & 0.73                    \\
Iraq                         & 1/6/2022                  & Crimean-Congo Hemorrhagic Fever (1940s)                         & 13.40\%                      & 0.09                    \\
African region               & 10/6/2022                 & Monkeypox (1970s)                                               & 4.69\%                       & 2.57                    \\
Pakistan                     & 17/6/2022                 & Cholera (19th century)                                                 & 0.70\%                       & 0.07                    \\
Somalia                      & 20/7/2022                 & Cholera (19th century)                                               & 0.47\%                       & 0.19                    \\
Ghana                        & 22/7/2022                 & Marburg virus disease (1967)                                   & 100.00\%                     & 2.00                    \\
Bangladesh                   & 28/11/2022                & Dengue (18th century)                                              & 0.43\%                       & 0.71                    \\
Tanzania                     & 12/8/2022                 & Leptospirosis (18th century)                                          & 20.00\%                      & 0.12                    \\
African region               & 3/1/2023                  & Yellow fever (17th century)                                           & 8.79\%                       & 0.12                    \\
Argentina                    & 5/9/2022                  & Legionellosis (1977)                                           & 36.35\%                      & 0.25                    \\
Uganda                       & 8/12/2022                 & Ebola disease caused by Sudan virus (1976)                    & 38.73\%                      & 0.72                    \\
Nepal                        & 10/10/2022                & Dengue (18th century)                                                  & 0.14\%                       & 0.14                    \\
Haiti                        & 13/12/2022                & Cholera (19th century)                                                & 2.07\%                       & 4.35                    \\
Pakistan                     & 13/10/2022                & Dengue (18th century)                                                  & 0.24\%                       & 0.23                    \\
Pakistan                     & 17/10/2022                & Malaria (Paleogene period)                                                 & --                           & --                      \\
Lebanon                      & 19/10/2022                & Cholera (19th century)                                                 & 11.11\%                      & 0.29                    \\
Mauritania                   & 20/10/2022                & Rift Valley fever (early 1900s)                                     & 48.94\%                      & 0.48                    \\
Niger                        & 8/2/2023                  & Meningitis (19th century)                                             & 16.22\%                      & 0.21                    \\
Demo. Rep. of Congo & 10/2/2023                 & Cholera (19th century)                                                 & 0.36\%                       & 0.31                    \\
Bangladesh                   & 17/2/2023                 & Nipah virus infection (1999)                                  & 72.73\%                      & 0.20                    \\
Mozambique                   & 24/2/2023                 & Cholera (19th century)                                               & 0.71\%                       & 0.23                    \\
Equatorial Guinea            & 25/2/2023                 & Marburg virus disease (1967)                                  & 96.55\%                      & 0.38                   \\
\hline
\end{tabular}
\end{table}

Therefore, as in the future new pandemics are going to spread due to new issues like global warming (see, for example, \cite{Christie2021} for possible new dangerous and unknown viruses generated from the thawing of permafrost in Siberia), new intensive farming and cultivation techniques (see, for example, \cite{Moreno2023} on how high population density in farming single animal species may integrate with other factors to increase the risk of mutation, re-assortment, and the generation of new pathogens) and increased migration, urbanization and the number of wars at a global level (see, for example, \cite{Hoiby2020}), and probably with characteristics similar to that of COVID-19 in terms of speed of spreading, contagion rates and impact on societies, the pandemic "data challenge" should be one of the most important global tasks we must focus on. It is not a matter of chance that during the COVID-19 pandemic, among western countries, those most hit had problems with their data collecting systems. It is true that the viral characteristics of COVID-19 do not allow a clear country comparison with respect to the efficiency of facing the pandemic. In fact, \cite{Martinez2021} found that European and American countries were less efficient than South Asian and African countries, but they also admitted that this was mainly due to demographic features of the populations, hidden mortality in African and South Asian countries and the virus hitting mostly the elderly and fragile people. However, other viruses with similar contagion speeds and rates but hitting youngsters rather than the elderly might be hard to face if data collection is not efficient. 

In this paper, we consider the bias between excess mortality and the official Italian COVID-19 data in the first 2020 outbreak for evaluating data quality in a space-time context. To model this bias we opted to use a Bayesian framework where two different quality measures ought to be evaluated: (i)  the share in the population dying because of a particular infectious disease without being officially reported, so large values of this measure represent a worse scenario, and (ii) the coverage of the epidemic by the health systems, which can be considered an adequate indicator of their quality and a proxy for the efficacy of the crisis response. Among the factors explaining the bias variability, the focus is on detecting the most important component among spatial, temporal and interaction components.

The paper is organized as follows. Section 2 is devoted to a description of the data used, Section 3 focuses on proposed metrics for quantifying bias in official death data. In Section 4 we present the spatial, temporal, and spatial-temporal Bayesian model 
%proposed by Franco-Villoria, \cite{FrancoVilloria2022} 
for the proposed metrics. The results are displayed in Section 5. Finally, Section 6 concludes the paper with an overview of possible future work.
\section{Data}
In order to evaluate the data quality on epidemic mortality in general and on COVID-19 mortality in particular, we considered two data sources: (i) official data on the pandemic evolution considering the essential variables to be considered for classic SE(I)RD models (daily or weekly new infections  - i.e. cases -, susceptible, exposed, recovered and deceased people). More often similar data collection starts \textit{ad hoc} at the request of national and international organisms to face the epidemic as was the case for the COVID-19 pandemic. In the following, we refer to this data as \textit{official data}. (ii) National or supranational statistical institute data on population mortality. Henceforth, we refer to this data as \textit{ISTAT data} as our main application will be on the Italian case and ISTAT is the Italian national statistical agency.

\paragraph{Official data}
In Italy, official data about deaths related to COVID-19 was reported daily from the 24th of February 2020 to the 30th of October 2022, at the European Union NUTS-2 level (i.e. regions), by the Italian Ministry of Health. While information about new cases was also published at the more refined NUTS-3 level (i.e. provinces), the number of daily COVID-19 new deaths was not officially available at this level. Nevertheless, it was possible to reconstruct the time series of COVID-19 at a NUTS-3 level indirectly, using other official sources like regional authorities' daily bulletins on provincial new cases, hospitalization and deaths and other information sources \cite{Ferrari2021}. Bulletins were in general in a "pdf" format so we were able to scrape data from these documents and to retrieve the data of interest for the majority of the Italian provinces.
%, however lacking some of the provinces in the Lombardy and Campania regions - Lombardy being undoubtedly the worst hit area in the first wave of COVID-19 in Italy. 

In March 2022, ISTAT published data from Istituto Superiore di Sanità (ISS - the main public health institute in Italy) about the monthly evolution of COVID-19 deaths in each province. Using the regional weekly trends, it was possible to reconstruct an estimate for the weekly provincial COVID-19 officially reported deaths for the provinces which were missing from the scraped data. This technique has also been applied for some provinces for which the data had been scraped but the discrepancy with this monthly report was significant for some provinces, especially those experiencing low levels of mortality.

Official data about deaths caused by COVID-19 is assumed to have been subject to delays, errors, inconsistencies between reporting protocols, etc. Moreover, it is sensible to assume that the data has a systematic underestimation and is therefore biased as an estimate for actual COVID-19-related deaths: especially in the first pandemic stages, testing and care were not available for all people infected with the disease, and the official data only reflected the share of people that went through at least a minimal contact with the health system \cite{Castaldi2020, Castaldi2021, Rivieccio2021}. Of course, this bias is supposed to change over time, as well as space, and in general, it is assumed to be directly proportional to the severity of the epidemic.

Additionally, COVID-19 did not only cause deaths directly because of infection, but the burden on the health system caused by the epidemic prevented to cure other diseases and accidents, causing indirect deaths of other individuals as well. Although lockdown measures might have prevented some of these "traditional" causes of death, this aspect should nevertheless be taken into account when assessing the impact on mortality caused by COVID-19.

With respect to the period under consider
%?? add something about the cut of the first wave because we need to cut and do 2 waves because there is a change in the number of provinces available? so we analyze the first wave and then the second with a subset of provinces??

\paragraph{ISTAT data}
Each year, ISTAT provides a daily record of deaths reported in each municipality of Italy \footnote{Source: \url{https://www.istat.it/en/archivio/268504}}. In this study, the ISTAT data is aggregated by province and weekly: this is because the corresponding COVID-19 official data has a strong seasonality over the weekdays. This detailed time series data can help estimate the actual deaths in 2020 caused by COVID-19, both directly and indirectly. 
%%%%%%%%%%%%%%%%%%%%%%%%%%%%%%%%%%%%%%%%%%%%%%
A straightforward way to estimate this is through the concept of "excess mortality", as the excess between the 2020 overall mortality and the average in the previous few years.
%%%%%%%%%%%%%%%%%%%%%%%%%%%%%%%%%%%%%%%%%
 
In this work, it is proposed to use a 5-year window consisting of the period 2015-2019 to represent the stable mortality level. The excess mortality is then found by subtracting this stable level from the 2020 deaths data, in each province and week of the year. While the 2015-2019 average will be smoother, the 2020 time series is subject to a great amount of noise, so it is chosen to apply minimal smoothing, consisting of a 3-week moving average, to make it more stable. 

%In order to make this data homogeneous with the official COVID-19 data, the first ??? weeks of 2020 were removed from the series. Let 
%?? add something about the cut of the first wave because we need to cut and do 2 waves because there is a change in the numebr of provinces available? so we analyze the first wave and then the second with a subset of provinces??

Since no other relevant information is available, this excess mortality can be imputed to COVID-19 and be an estimate for the actual number of deaths caused by COVID-19, both directly and indirectly. 
Indeed, in many works the indirect contribution of the excess mortality due to COVID-19 is highlighted, see, for example, \cite{Dorrucci2020}, \cite{Achilleos2020}, \cite{Modig2021} and \cite{Vanella2021}.
Mixing this estimate for the actual number of deaths with the official data allows for creating measures for the under-reporting bias present in the official figures. Such metrics can be interpreted as a proxy for the quality of the health system response to the crisis and to assess how this evolved and changed over time and space. In the end, the spatial distribution of these metrics can be extremely useful for the policy-maker to identify hotspots from the health system network, with the best and worst response to the emergency. Finally, studying the local policies and procedures implemented in those highlighted areas can help in the definition of best practices for future emergencies and epidemics.

\section{Proposed metrics}
The aim of the metrics to be defined is to provide an estimate for the under-reporting mortality bias that the official data has been subject to. However, there is no unique way to define such bias. Here, two different metrics with different interpretations for the policy-maker are proposed.

\subsection{Additive bias $b^{A}$}
Let $D_{ij}$ be the officially reported total number of deaths in province $i$ and week $j$ which exceeds the average of the previous 5 years, assuming that they can all be imputed to the COVID-19 emergency. 
Let $\hat{D}_{ij}$ be an estimate for $D_{ij}$. Let $Y_{ij}$ be the officially reported number of COVID-19-related deaths in province $i$ and week $j$. 
Finally, let $POP_{i}$ be the average population in province $i$ along the considered period. The additive bias is built as the difference between the \textit{actual mortality} $D_{ij}/POP_{i}$ and the \textit{official mortality} $Y_{ij}/POP_{i}$. This bias is defined as "additive" because it must be added to the official mortality to get the unbiased value:

    \begin{gather*}
         \tilde{b}^{A}_{ij}=\frac{\hat{D}_{ij}-Y_{ij}}{POP_{i}}
    \end{gather*}
    
    Although, at least theoretically, we have that $D_{ij}\geq Y_{ij}$, $\tilde{b}^{A}_{ij}$ can take in practice negative values because $\hat{D}_{ij}$ can take values lower than $Y_{ij}$ and even be negative by design. Moreover, even assuming that $\hat{D}_{ij}$ correctly estimates ${D}_{ij}$, errors and delays potentially occurred in the reporting of $Y_{ij}$. Negative values of $\tilde{b}^{A}_{ij}$ can be removed and/or treated as 0.

    In terms of interpretation, $\tilde{b}^{(A)}$ defines the share in the population that died because of COVID-19 without being officially reported, so large values represent a negative scenario. Its trend over time and space (but not its magnitude) is a rough proxy for the part of the pandemic that was concealed and undetected by the public administration. Also, it can be roughly interpreted as the risk for an individual in a population of being infected and die of COVID-19 without having had the possibility to access the healthcare system because of capacity limits: this is of course dangerous for the population, since access to the healthcare system reduces the consequent risk of dying because of the disease. However, it is a personal risk because it does not consider the severity of the outbreak in each point in time and space, but it only assesses the remaining part of the pandemic that was unfortunately missed out by the competent authorities.

    For example, two populations with the same $b^{A}=K$ have the same individual risk of dying of COVID-19 outside of the healthcare system, but the system itself may have to deal with two very different spread of the disease in their population so that the same value of $b^{A}$ can be considered decent result in one setting but utterly unsatisfactory in others.
    
    Overall, it can be interpreted as the damage or impact of under-reporting on the population's well-being, but it is not an appropriate indicator for the quality of the healthcare system. In order to make this metric comparable with the following one, scaling is applied to the original formula:

    \begin{gather*}
         {b}^{A}_{ij}=1000\cdot \tilde{b}^{A}_{ij} =1000\cdot\frac{\hat{D}_{ij}-Y_{ij}}{POP_{i}}
    \end{gather*}

\subsection{Multiplicative bias $b^{M}$} 
    
    The ratio between $Y_{ij}$ and $D_{ij}$ assesses the probability of a COVID-19-related death being officially reported. In order to transform it into a bias metric, the complementary probability of not being reported is considered instead and called $b^{(M)}_{ij}$. This is equivalent to the additive bias divided by the excess mortality rate.

    \begin{gather*}
        b^{M}_{ij}=1-\frac{Y_{ij}}{\hat{D}_{ij}}= \frac{b^{A}_{ij}}{\hat{D}_{ij}/POP_{ij}}
    \end{gather*}

    This metric should also stay theoretically between 0 and 1, but for the same issues cited before, it can exceed these boundaries. Again, a larger bias indicates a bad situation. Regarding its meaning, $ b^{M}$ measures the coverage of the pandemic by the healthcare system, thus it is an adequate indicator of its quality and a proxy for the efficacy of the crisis response. 
    
    The quality of the response is assumed to have been at least partially correlated to the evolving severity of the pandemic, e.g., a good emergency policy might nevertheless have performed poorly at the peak of the epidemic, while an ill-advised response policy may have exceeded expectations because of the negligibility of the crisis it had to face. Hence, the distribution of the multiplicative bias over time and space shall be assessed both in the absence and presence of a covariate approximating the intensity of the pandemic.

\paragraph{Processed data}
In this study, we consider the period starting from the first day of COVID-19 official data release (24th February) until the 11th May, after 11 weeks, when the national lockdown was lifted in Italy. We opted to consider this first wave window, as underreporting and overall data quality are at their lowest at the beginning of an epidemic for obvious reasons, and they tend to improve over time. 

\section{Model}

Inspired by the work of Franco-Villoria et al. (2022), \cite{FrancoVilloria2022}, the model aims to consider the spatial, temporal and spatio-temporal structure (as well as potential fixed effects) for $b^{A}_{ij}$ and $b^{M}_{ij}$. The objectives of such models are: assessing the overall spatial distribution of these quality indicators on the Italian territory, as well as the temporal trend in the first wave of COVID-19; assessing the significance of the interaction structure in the model; estimating the importance of each of these components, including also the contribution of the potential fixed effects.

The chosen specification is a latent Gaussian model on the logit transformation of the response with random effects. The same modelling structure can be applied to both response variables. In matrix notation:

\begin{gather*}
\mathbf{b}=[b_{1,1},...,b_{1,j},...,b_{1,J},...,b_{i,1},...,b_{i,J},...,b_{I,1},...,b_{I,J}]^T\\
\text{logit}(\mathbf{b})|\boldsymbol{\eta},\sigma^2_\epsilon \sim N_{I\cdot J}(\boldsymbol{\eta},\sigma^2_\epsilon \mathbf{I})\\
\;\\
\boldsymbol{\eta}=\boldsymbol{\mu}+\boldsymbol{\sigma}_u(\mathbf{I}_n \otimes \mathbf{1}_n)\mathbf{u}+\boldsymbol{\sigma}_v(\mathbf{1}_n \otimes \mathbf{I}_m)\mathbf{v}+\boldsymbol{\sigma}_w\mathbf{w}\\
\;\\
\mathbf{u}=[u_1,...,u_J]^T\sim N_J(\mathbf{0},\boldsymbol{\Sigma}_u)\;\;\;\;\;\;\;\;\;
\mathbf{v}=[v_1,...,v_K]^T\sim N_K(\mathbf{0},\boldsymbol{\Sigma}_v)\\
\mathbf{w}=[w_{1,1},...,w_{1,K },...,w_{J,1},...,w_{J,K}]^T\sim N_{J\cdot K}(\mathbf{0},\boldsymbol{\Sigma}_w=\boldsymbol{\Sigma}_u\otimes \boldsymbol{\Sigma}_v)
\end{gather*}

where $\mu$ represents the  the mean value all times and spaces. $u_i$ and $v_j$ respectively represent the spatial and temporal main effects, and $w_{ij}$ is the interaction random effect. This structure allows us to clearly distinguish between the three main sources of variation under investigation, i.e. a spatial association, a changing pattern over time, and an interaction between these two elements.

The parameters of this model are $\mu,\boldsymbol{\beta},\sigma_u,\sigma_v,\sigma_w,\sigma_\epsilon$, while the covariance matrices $\boldsymbol{\Sigma}_u \text{ and }\boldsymbol{\Sigma}_v$ respectively define appropriate spatial and temporal models and are treated as fixed and known. The covariance matrix of the interaction term is defined as the Kronecker product of the covariance matrix of the two main effects, following the work of Knorr-Held, \cite{KnorrHeld2000}.

\paragraph{Temporal component specification\\}
The objective of the model is to retrieve the main temporal pattern behind the response and assess the impact of this factor. Hence, a $RW_1$ model could be a simple but flexible choice, appropriate for the goal of the project. Note that this is not assumed to be the actual temporal model behind the data but it is exclusively used because of its convenience.

\begin{gather*}
    \mathbf{v}\sim N_K(\mathbf{0},\boldsymbol{\Sigma}_v=\boldsymbol{\Sigma}_{RW1}).
\end{gather*}

\paragraph{Spatial component specification\\}
One of the most popular models for lattice data is the ICAR model \cite{Besag1974} \cite{Besag1991}, which consists of an improper Normal distribution for the spatial component, defined using a square symmetric matrix $\mathbf{M}$ with only non-negative entries and a null diagonal, and a corresponding diagonal matrix $\mathbf{D}$ where $d_{p,q}=\sum_q{m_{p,q}}$.

\begin{gather}
  \mathbf{u} \sim N_J \left(\mathbf{0};\boldsymbol{\Sigma}_u=(\mathbf{I_J-D^{-1}M})^{-1}  \mathbf{D}^{-1}\right).
\end{gather}\\
The weights $m_{p,q}$ represent the neighbourhood structure between the spatial areas, and it is usually based on the adjacency matrix, according to which regions that share a border are assigned $m_{p,q}=1$ and 0 otherwise. This approach is appropriate when the adjacency is the most reliable factor providing information about the potential association between regions. However, it can be argued that other variables should be considered instead, whenever they are more reasonable estimators of the actual connection between different areas. Adjacency matrices often consider geographical factors inappropriately: e.g. islands are isolated from the mainland, while geographical barriers such as mountains are ignored. 

This is particularly relevant in the context of epidemiology, where the main factor causing spatial association is human mobility. This is considered the prevalent factor behind spatial association so traditional definitions of the weight matrix $M$, such as adjacency or distance-based matrices are unable to capture factors such as commuting routes, physical boundaries, air and highways traffic, metropolitan areas, etc.

Hence, it is proposed to replace the traditional adjacency matrix with smartphone location data, which actually estimates the average commuting of individuals between two provinces, no matter their actual geographical location.   The spatial weight matrix used in this application was derived starting from data provided by Pepe et al. \cite{Pepe2020}, who built a daily time series of an origin-destination flow matrix processing smartphones' location data across Italy. The daily matrices from 18th January to 21st February were averaged to create a mean matrix representing the average mobility across provinces before the beginning of the COVID-19 outbreak. The matrices were built so as to have rows summing to 1, including the within province mobility. In order to respect the conditions necessary for the CAR model, the diagonal entries were set to 0 and the remaining entries were normalised through a division by row sum. The origin-destination matrix was not symmetric as the flows were directional: hence, the inward and outward flow between two provinces were averaged in order to obtain an appropriate $\mathbf{M}$.

\begin{figure}[H]
    \centering
    \includegraphics[scale=0.75]{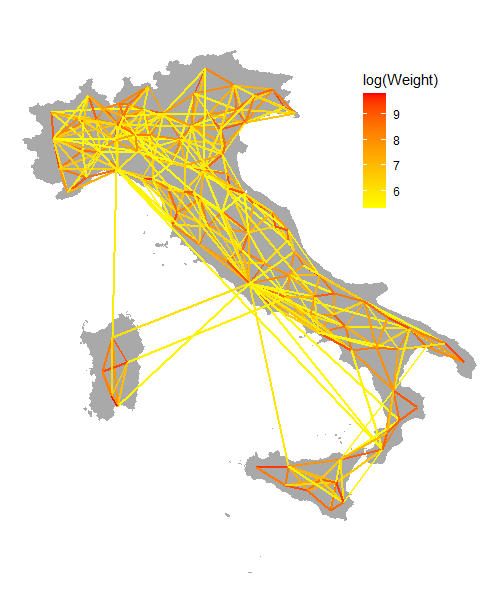}
    \caption{Mobility-based weight matrix in log scale}
    \label{fig:spatial_weights}
\end{figure}

The single disadvantage of this approach consists of the loss of the sparsity of the matrix. In order to obtain a matrix sparse enough for computation efficiency, it was chosen to consider only the last quintile, i.e. the highest 20 \% of the weights, assuming irrelevant the impact of smaller associations. The resulting weights can be visualized in Figure \ref{fig:spatial_weights}: it is clear how this weight matrix is an improvement with respect to the naive adjacency one, as it is able to recognize important industrial hubs and big cities, as well as air and sea traffic, giving an overall accurate representation of the spatial structure.

\begin{figure}[H]
    \centering
    \includegraphics[scale=0.75]{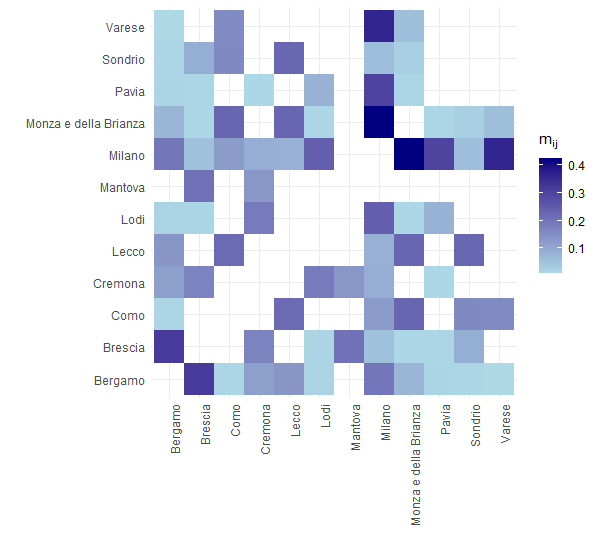}
    \caption{Weights $m_{ij}$ between the provinces of Lombardy}
    \label{fig:lombardy_spatial}
\end{figure}

A snippet of the $\mathbf{M}$ for the provinces of Lombardy is represented in Figure \ref{fig:lombardy_spatial}: this plot shows how the smartphone location data precisely reconstruct the human mobility patterns not only at the national level but also locally.

\subsection{Prior specification}
In terms of prior specification, the same approach of Franco-Villoria et al. \cite{FrancoVilloria2022}, as the innovative hierarchical variance decomposition method introduced by Fuglstad et al. \cite{Fuglstad2020} is employed in the context of spatio-temporal epidemiological data. The method is based on a reparametrization of the variance parameters in terms of a single total variance and a set of proportions, defined through a decomposition tree. In this setting, it becomes easier to translate prior assumptions on the parameters into prior distributions and hyperparameters. Moreover, it gives more intuitive results in the posterior analysis, as the parameters are already set as adimensional proportional contributions, rather than variances or standard deviations.

As in Franco-Villoria et al. \cite{FrancoVilloria2022}, here we choose to define the first split to separate the main from the interaction term, and secondly, the spatial and temporal effects are divided. This results in a new reparametrization of the three original variances $\sigma^2_ u,\sigma^2_v,\sigma^2_w$ into a total residual variance $V$, the proportion $\psi$ of this $V$ given by the interaction term, and the proportion $\phi$ of main effects variance imputable to the spatial effect. The prior specification is then chosen on this new set of parameters. Specifically, the INLA default prior on variance parameters on $\sigma^2_\epsilon$, a Uniform on $\phi$, and a Penalized Complexity (PC) prior \cite{Simpson2017} on $\psi$ with base model $\phi_0=0$, and a PC prior on $V$ with base model $V_0=0.$

\begin{gather*}
\sigma_u^2,\sigma_v^2,\sigma_w^2 \longrightarrow V,\phi,\psi \\
\boldsymbol{\eta}= \sqrt{V}\left\{\sqrt{1-\psi}\left[\sqrt{1-\phi}(\mathbf{I}_n \otimes \mathbf{1}_n)\mathbf{u}+\sqrt{\phi}(\mathbf{1}_n \otimes \mathbf{I}_m)\mathbf{v}\right]+\sqrt{\psi}\mathbf{w}\right\}\\
V \sim PC_0(U,\alpha=0.05)\\
\phi \sim Unif(0,1)\\
\psi \sim PC_0(\lambda_\psi=1)\\
\sigma^2_\epsilon \sim Gamma(1,5e^{-5}).
\end{gather*}

\section{Results}
\subsection{Fitting the models for $b_A$ and  $b_M$}
The models were fitted in R using the INLA software \cite{INLA}. The same model was fitted for $b_A$ and $b_M$, with a difference in the prior specification in the $U$ upper bound parameter of the PC prior on $\psi$, to reflect the different variability levels found in the two responses. Specifically, $U_A=0.1$ and $U_M=1$. More details can be found in \cite{Simpson2017}.

\begin{figure}[H]
    \centering
    \includegraphics[scale=0.5]{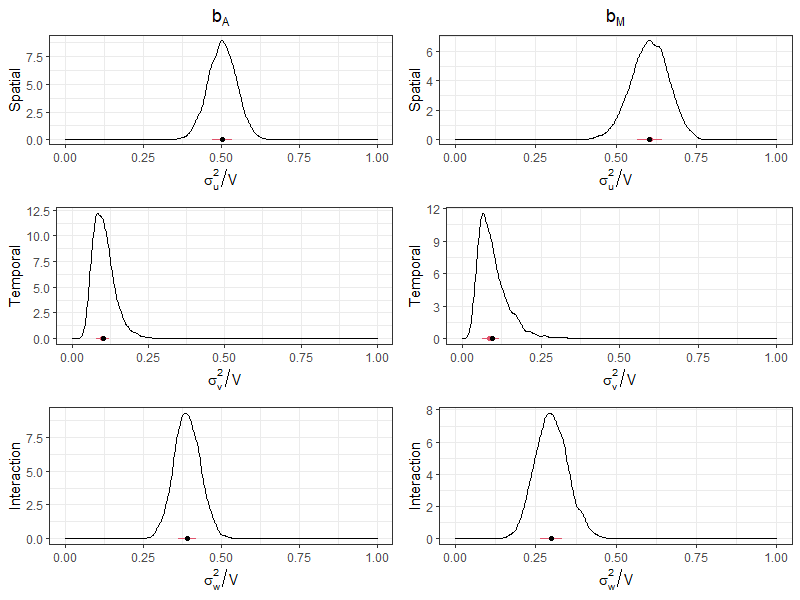}
     \caption{Posterior densities of the variance parameters as proportions of the total residual variance $V$ in the $b_A$ and $b_M$ models}
      \label{fig:proportions}
\end{figure}

First of all, the different contributions of the random components to the total variability can be assessed considering the marginal posterior distribution of proportions of total variance $V$, see Figure \ref{fig:proportions}. For both $b_A$ and $b_M$, it appears that the spatial component is the most relevant, followed by the interaction effect, while the temporal trend explains a much smaller share of the total variability. Specifically, the spatial component explains around 50 \% of the structured variability of $b_A$, while around 60 \% for $b_M$. This first result already shows how the multiplicative bias $b_M$ is more influenced by the spatial structure than $b_A$, which means it is more correlated with the intrinsic characteristics of the local areas, e.g. health system quality.

\begin{figure}[H]
    \centering
     \includegraphics[scale=0.75]
    {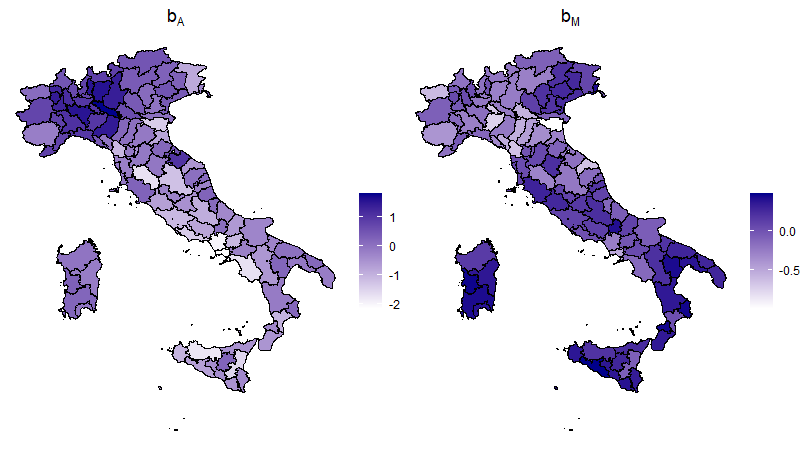}
      \caption{Posterior mean of the spatial random effects on $b_A$ and $b_M$}
      \label{fig:spatial_effect}
   
\end{figure}

Secondly, the posterior means of the random effects offer a summary of the different contributions to the response. Figure \ref{fig:spatial_effect} shows the posterior mean of the spatial random effects over the provinces of Italy. For 
$b_A$ on the left, provinces in the North of Italy experienced a larger share of underreported deaths with respect to the overall population. However, the spatial distribution completely changes for the $b_M$ metric, as most of the Northern provinces show small values, while the highest effects are found in the Southern and North-Eastern provinces. These figures display how the two indices measure very different quantities, with $b_M$ being much more consistent with the literature on the spatial distribution of health system quality indicators in Italy.

\begin{figure}[H]
    \centering
    \includegraphics[scale=0.65]{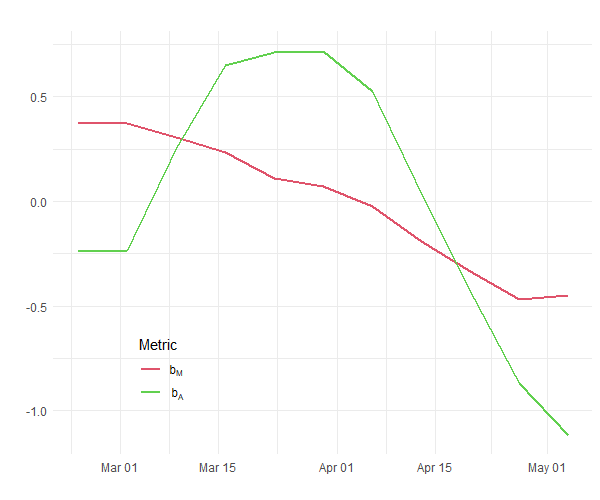}
    \caption{Posterior mean of the temporal random effect on $b_A$ and $b_M$} \label{fig:temporal_effect}
   
\end{figure}

With respect to the temporal pattern, the two metrics also show differences. The average temporal trend for $b_A$, shown in Figure \ref{fig:temporal_effect}, green curve, starts with an increasing part, up to the sixth week in the considered period, followed by a steady fall in the remaining weeks. This is a reasonable result, as it is expected that the indicator $b_A$ performed the worst at the peak of the "official" epidemic evolution, plus a delay due to the fact that deaths are considered instead of cases. Hence, this confirms the assumption that $b_A$ is related to the level of stress of the health system, rather than to the quality of its response to a certain amount of stress.\\
The results for $b_M$ are again completely different as the posterior means, shown in Figure \ref{fig:temporal_effect}, red curve, %of the temporal effects 
show a steadily decreasing trend. It is important to note that the level is at its maximum at the beginning of the data reporting period and then, the quality level of the emergency response increases as the COVID-19 situation is taken more and more seriously and more effective policies and practices are put in place. Of course, this shows how the official data are unreliable, not only in magnitude but also in their trend, and should not be used raw to evaluate the evolution of an epidemic, particularly at the beginning of the reporting period, as the quality of the official data tends to significantly improve, i.e. a decrease in the $b_M$ indicator. 

\begin{figure}[H]
    \centering
 \includegraphics[scale=0.5]{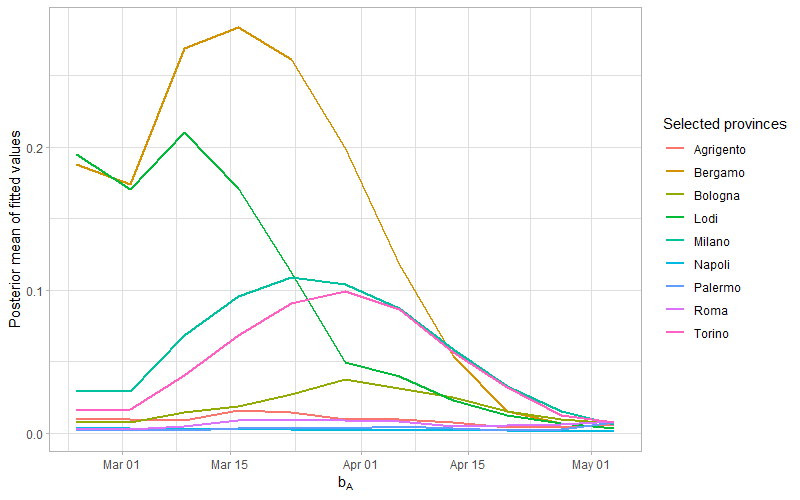} \caption{Posterior mean of the fitted values for $b_A$ in the original scale for selected provinces}
  \label{fig:b_a_selected_provinces}

\end{figure}

\begin{figure}[H]
    \centering
    \includegraphics[scale=0.5]{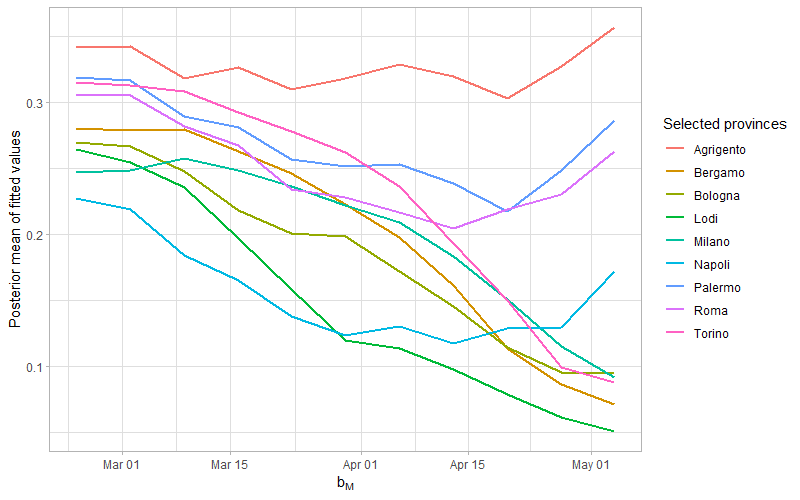}
     \caption{Posterior mean of the fitted values for $b_M$ in the original scale for selected provinces}
      \label{fig:b_m_selected_provinces}
\end{figure}

While the temporal and spatial effects can offer general evaluations of the overall evolution, the interaction factor seems to play a vital role in the total variability, as shown in Figure \ref{fig:proportions}. Hence, any conclusion on the single provinces' behaviour should be made on the whole linear predictor. In the following figure, we considered the posterior distribution of the fitted values reparametrized in the original scale of the $b$ metrics, through a logistic transformation. The provinces were selected among the most populous areas in Italy, along with provinces particularly affected by COVID-19, and provinces displaying surprising behaviours in the results. The importance of the interaction term appears clearly in Figures \ref{fig:b_a_selected_provinces} and \ref{fig:b_m_selected_provinces} and, as the provinces display quite a variety of temporal evolution behaviours.  

\subsection{Clustering for the $b_M$ indicator}

\begin{figure}[H]
    \centering
    \includegraphics[scale=0.75]{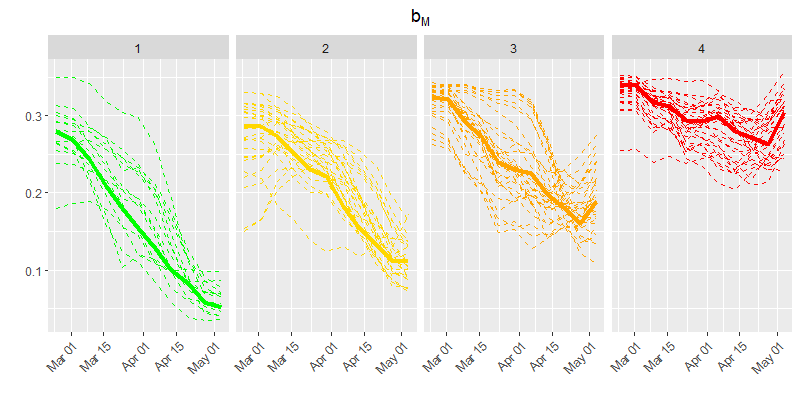}
    \caption{Posterior mean of the fitted values divided in 4 clusters with corresponding centroids in the provinces of Aosta, Rimini, Catanzaro, and Cosenza}
    \label{fig:b_m_cluster_time_series}
\end{figure}

In order to classify provinces on the basis of the 
$b^M$ metric, i.e. a data quality indicator, clustering was applied to the posterior means of the fitted values from Model $b_M$. Specifically, partitional clustering using a dynamic time warping distance was chosen as an appropriate partitioning method for time series \cite{tsclust}. The posterior mean of the fitted values from the model were chosen as the time series, in order to remove the residual noise present in the original raw data. Evaluating different performance metrics, the optimal cluster number was found to be 4. Figure \ref{fig:b_m_cluster_time_series} shows the 4 different classes and their centroid, ordered by best to worst overall performance. 

The first cluster, coloured in green, includes all the provinces with a steady and relevant decrease over time; moreover, all of them except one also even start from a relatively low level already at the beginning of the reporting period. The red and last cluster identifies the worst-behaved class, as the index $b_M$ is already quite high at the beginning of the period is mostly stable, and does not display significant signs of improvement during the first wave.

The other two clusters represent intermediate behaviours and they are more ambiguous and heterogeneous. First, the orange cluster identifies provinces that performed better than the ones in the red cluster, but worse than the green provinces, because of a much lower rate of decrease. There are two main subtypes of trends in this cluster. Most of the provinces display an initial decrease, followed by a mild rise in the last few weeks, which could be a sign of a decline in alertness and a possible relaxation of some emergency measures initially put in place. Few provinces in the same cluster actually followed an inverse pattern, with an initial stable level of $b_M$ followed a quite steep improvement (e.g. Udine, Pordenone, Verona, Belluno, Vicenza, Padova, Venezia, Treviso), as if the severity of the epidemic was downplayed after the initial shock. The two groups have in common an insufficient degree of vigilance and promptness in dealing with an epidemic emergency, which became manifest as either a delay in putting in place effective response policies or a relaxation of the same way too early.

For the second group, the general trend, followed by the majority of the units in this cluster, is decreasing but more slowly than for the green provinces, hence indicating a worse performance in data reporting. However, looking more closely, we can identify a small subgroup of provinces in this cluster that displays an unusual concave U-shape trend: Sondrio, Verbano-Cusio-Ossola, Monza e della Brianza, Varese, Lecco, Como. The areas with this distinct behaviour are all located in North-West Italy and all fall into this class because they do not tend to reach extremes, either positive or negative, but are more or less stable around the 0.2 threshold. One hypothesis could be that these provinces might have experienced an actual delay in the spread of the epidemic with respect to the general evolution, which might have led the peak to be shifted later than the other province, for about 5 or 6 weeks. This delay was probably responsible for keeping $b_M$ low even in the worst moments for these provinces, as they had a temporal advantage consisting of all the experience and knowledge accumulated and shared by the other health system administrations over the first few weeks.

\begin{figure}[H]
    \centering
    \includegraphics[scale=0.75]{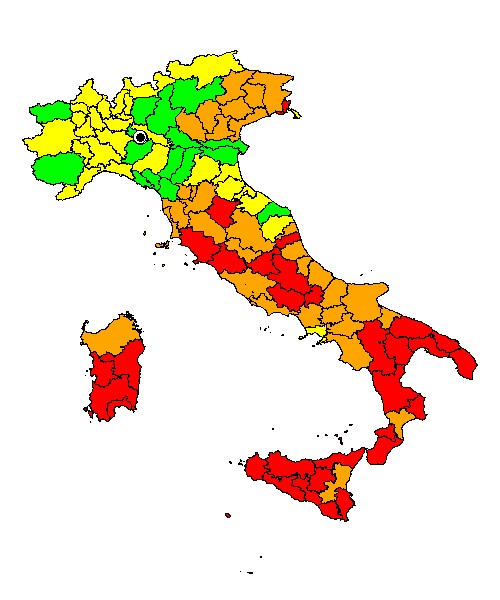}
    \caption{Provinces by cluster and the city of Codogno (LO) marked by a black dot}
     \label{fig:b_m_clustering_map}
\end{figure}

The map presented in Figure \ref{fig:b_m_clustering_map} summarizes the cluster assignments over the Italian provinces, along with the location of the first COVID-19 hotspot in Italy in Codogno, Lodi. At first glance, it may seem counter-intuitive that areas closer to the first outbreak performed better than the ones far away. However, it may be that provinces in the North considered the emergency more seriously and sooner than the rest of Italy, as they perceived the risks earlier and may have put in place effective response protocols before the Southern areas.

\section{Conclusions and future work}
In this study, we considered two metrics to evaluate the quality of epidemic official data. This has been achieved using a measure of excess mortality registered during an epidemic period, with respect to the average level in the previous years. While the metric $b_A$ is a proxy for the underestimation of the epidemic severity, it has been shown how $b_M$ can be used as a quality indicator of the health system in terms of monitoring the epidemic situation and reporting accurately the data. Both metrics were smoothed through a spatio-temporal model, which highlighted the importance of the spatial components in the variability of both responses. Results for the $b_M$ are consistent with the assumptions about the data quality: in particular, the general temporal trend shows a steady decrease and it is in line with an improvement in the data quality in the first period of reporting. 

Finally, the Italian provinces were grouped on the basis of the $b_M$ indicator fitted values, in order to classify them into clusters of different quality levels. This could be extremely useful for policymakers, since it offers an overview of the reporting performance of the different health systems and may suggest strategies employed locally, which may become best practices for future epidemics.

Future work will entail an implementation of the quality metrics to other epidemics and nations. We may be interested in comparing the overall data quality between nations facing the same epidemic, rather than within a single nation. With respect to the Italian case, the time series could be extended to the following waves if official data about COVID-19 were available at the provincial level. Covariates related to the health system could be included in the models as fixed effects, to check whether they reduce the spatial variability and may explain the $b_M$ metric. 

In conclusion, the multiplicative bias is the main proposal of this work and it consists of a simple metric that could in theory be computed with a really short delay, i.e. as soon as registered deaths are aggregated by the governmental institutes for official statistics. Monitoring this indicator could help identify which areas display better results and performance in terms of the response to the epidemic, as well as areas where the severity of the situation is underestimated. This type of information could significantly accelerate the process of identification of effective and ineffective protocols and prevention measures, which consequently may save many lives down the line.

%\color{red}{ Another interesting aspect to investigate is the study of excess mortality according to different causes. Applying a decomposition of the excess mortality according to different causes could show which causes were attenuated and which causes were most inflated during the pandemic period. }

%\section{References}

% Use the following syntax and markup for your references if 
% the subject of your book is from the field 
% "Mathematics, Physics, Statistics, Computer Science"
%
% Contribution 

\end{document}